\documentclass{moriond}

\def\be{\begin{equation}}
\def\ee{\end{equation}}
\def\bea{\begin{eqnarray}}
\def\eea{\end{eqnarray}}

\usepackage{amsmath}
\usepackage{mathrsfs} 
\usepackage{amsfonts}
\usepackage{amssymb}
\usepackage{amsthm}
\usepackage{mathtools}
\usepackage{graphicx}
\usepackage{latexsym}
\usepackage{xspace}
\usepackage{hyperref} 
\usepackage{bm}
\usepackage{relsize}
\usepackage{tabularx}
\usepackage{multirow}
\usepackage{amssymb}
\usepackage{lscape}
\usepackage[most]{tcolorbox}
\usepackage{braket}
\usepackage{comment}
\usepackage[utf8]{inputenc}
\usepackage[english]{babel}
\usepackage{slashed}
\usepackage{empheq}
\usepackage{upgreek}
\usepackage{wrapfig}

\def\pia{\pi_{a}}
\def\pir{\pi_{r}}
\def \pia{ \pi_{a}}
\def \pir{ \pi_{r}}

\def\bmx{{\boldsymbol{x}}}

\def\bmk{{\boldsymbol{k}}}

\newcommand{\dd}{\mathrm{d}}

\newcommand{\Ima}{\Im \mathrm{m}\,}

\newcommand{\Eq}[1]{Eq.~(\ref{#1})}

\newcommand{\Fig}[1]{Fig.~{\ref{#1}}}

\newcommand{\Photo}{}

\begin{document}
\vspace*{4cm}
\title{Open Effective Field Theories for cosmology}

\author{ T. Colas }

\address{Department of Applied Mathematics and Theoretical Physics,\\ University of Cambridge, Wilberforce Road, Cambridge, CB3 0WA, UK}

\maketitle\abstracts{
When energy is not conserved, imprints of new physics on observable cosmology might not follow the rules of local effective actions. By capturing dissipative and diffusive effects, open effective field theories account for the possibly non-Hamiltonian evolution of cosmological inhomogeneities interacting with an unspecified environment. In this proceeding, we briefly discuss recent progress made towards their implementation in primordial cosmology. Our approach recovers the usual effective field theory of inflation in a certain limit and extends it to account for local dissipation and noises. Non-Gaussianities are generated that peak in the equilateral configuration for large dissipation and in the folded configurations for small dissipation. The construction provides an embedding for local dissipative models of inflation and a framework to study quantum information aspects of the inflationary models.}

\section{Introduction}

Organising a dialogue between the $\Lambda$CDM model and new physics is a central question to modern cosmology. For early universe physics, measurements of the temperature and polarization  anisotropies of the Cosmic Microwave Background (CMB) \cite{Aghanim:2018eyx} and the galaxy clustering of the Large Scale Structures (LSS) \cite{Colas:2019ret} are so far fully consistent with single-field slow-roll inflation. Simultaneously, the lack of detection of primordial gravitational waves leaves open the possibility that inflation took place at energy scales as high as $10^{14}~\mathrm{GeV}$, raising the hope of exploiting this cosmological era as a laboratory for new fundamental physics. Tremendous efforts have been made towards searching for new degrees of freedom, understanding of Quantum Field Theory (QFT) in curved spacetime, General Relativity (GR) at high energy or even having a glimpse of Quantum Gravity (QG). This hope is sustained by the general idea that, by gathering more and more data, we eventually become more sensible to the interplay between the visible and (so far) hidden sectors. 

In the context of inflation, CMB and LSS data are consistent with the existence of a single-scalar adiabatic degree of freedom in the early universe, the curvature perturbations $\zeta$, observed on scales from tens to thousands of MegaParsecs (Mpc). However, we suspect the presence of many other degrees of freedom during the early universe. Some are highly expected such as tensor modes, others are more speculative such as multifield constructions and high energy extensions. These extra degrees of freedom have not yet been directly detected and are so far only constrained. To account for their effects, the traditional approach consists in writing down a specific model on which we perform a set of perturbative computations. It leads to extract the summary statistics of $\zeta$ on large scales. These cosmological correlators can then be related to the late-time probes such as the temparature anisotropies of the CMB and compared to data. This approach has been successful in deriving the phenomenology one can expect from a new particle with a given mass, a given spin or a given set of interactions. The drawbacks are a proliferation of models and the lack of generic parametrisation to confront to the data.

\paragraph{Effective Field Theories}

Effective Field Theories (EFTs) provide a complementary perspective on this problem. Focusing on the (so-far) observed sector, they aim at incorporating unknown physics in a parametrically controlled manner. In this way, EFTs tame theoretical uncertainties related to the underlying microphysical description. Exploiting scale hierarchies and symmetries, EFTs focus on observational degrees of freedom, hereafter called \textit{the system}, while integrating out unobserved sectors (\textit{the environment}) and coarse graining over unobserved scales. In cosmology, EFTs such as the EFT of Inflation \cite{Cheung:2007st}, the EFT of the Large Scale Structures \cite{Carrasco:2012cv} and the EFT of Dark Energy \cite{Gubitosi:2012hu} have been instrumental to offer convenient parametrizations for data analysis. 

Recent years have shown through the development of the \textit{cosmological bootstrap} program \cite{Baumann:2022jpr} that physical principles tightly constrain the range of available dynamics for the system, independently of the details of the unknown environment. Soft theorems \cite{Hui:2022dnm} and near scale invariance provide a powerful organising principle. Combined with locality (Manifestly Local Test \cite{Jazayeri:2021fvk}) and unitarity of the IR description (Cosmological Optical Theorem \cite{Goodhew:2020hob}), there is little space remaining for the inclusion of new physics into this description. Obviously, these results only hold up to the validity of the above mentioned principles. Open EFTs precisely aim at relaxing the unitarity of the IR evolution for reasons we now develop.

\paragraph{Closed \textit{vs.} open} We consider the effective dynamics of a physical system. The closed/unitary perspective consists in accounting for the presence of a surrounding unknown environment by the inclusion of effective operators in the action of the system. By writing down all operators compatible with the symmetries of the problem and exploiting the scale hierarchies, we only keep a finite number of them, maintaining the predictive power of the theory. This construction encodes order by order the contributions of new physics into the Hamiltonian evolution of the system. Well-known examples are the iconic Fermi$-4$ theory or the field theoretic description of perfect fluids \cite{Dubovsky:2011sj}. This approach is extremely compelling and has demonstrated its versatility in an incredibly large amount of situations. Yet, not all situations in Nature can be described through this framework. Open systems require the inclusion of dissipative and diffusive effects which cannot be accounted for by a local effective action \cite{kamenev_2011}. These effects model the energy and information gains and losses of the system in contact with its environment. In the context of hydrodynamics, imperfect fluids experience viscosity which calls for new approaches \cite{Crossley:2015evo}, many of which pioneered the open EFT techniques presented in this proceeding. 

\paragraph{Cosmological Open Systems} Why should we bother about extending EFT constructions? One of the main differences between particle physics and cosmology is the lack of energy conservation in the latter case. The time-dependent homogeneous and isotropic background spontaneously breaks time-translation symmetry such that one does not expect the occurence of well segregated energy sectors, one for the UV and one for the IR. Dissipative effects and viscosity already play a crucial role in various cosmological phenomena such as reheating, Big Bang Nucleosynthesis or the EFTofLSS. 

To diagnose the need for inclusion of dissipative effects, vast efforts have been made in assessing entropy measures such as the \textit{purity} or the \textit{entanglement entropy} \cite{Colas:2022hlq}. 
While these quantities are kept constant under Hamiltonian evolutions, dissipative and diffusive dynamics generate information flows between system and environment. It appears that particle production out of the primordial vacuum favors these information exchanges between light species \cite{Colas:2021llj}. For instance, if one wants to model the evolution of the curvature perturbations while integrating out the tensor modes predicted by GR, one has to account for the non-Hamiltonian evolution of the curvature perturbations due to the non-linear interactions with the tensor sector \cite{Burgess:2022nwu}. Besides particle production related to the time-dependent background, many other frameworks lead to entropy generation such as instabilities, non-adiabatic features or warm inflation. Dissipative dynamics are then a common place in cosmology. 

\paragraph{Open Effective Field Theories}

\begin{wrapfigure}{r}{0.358\textwidth}
	\includegraphics[width=0.9\linewidth]{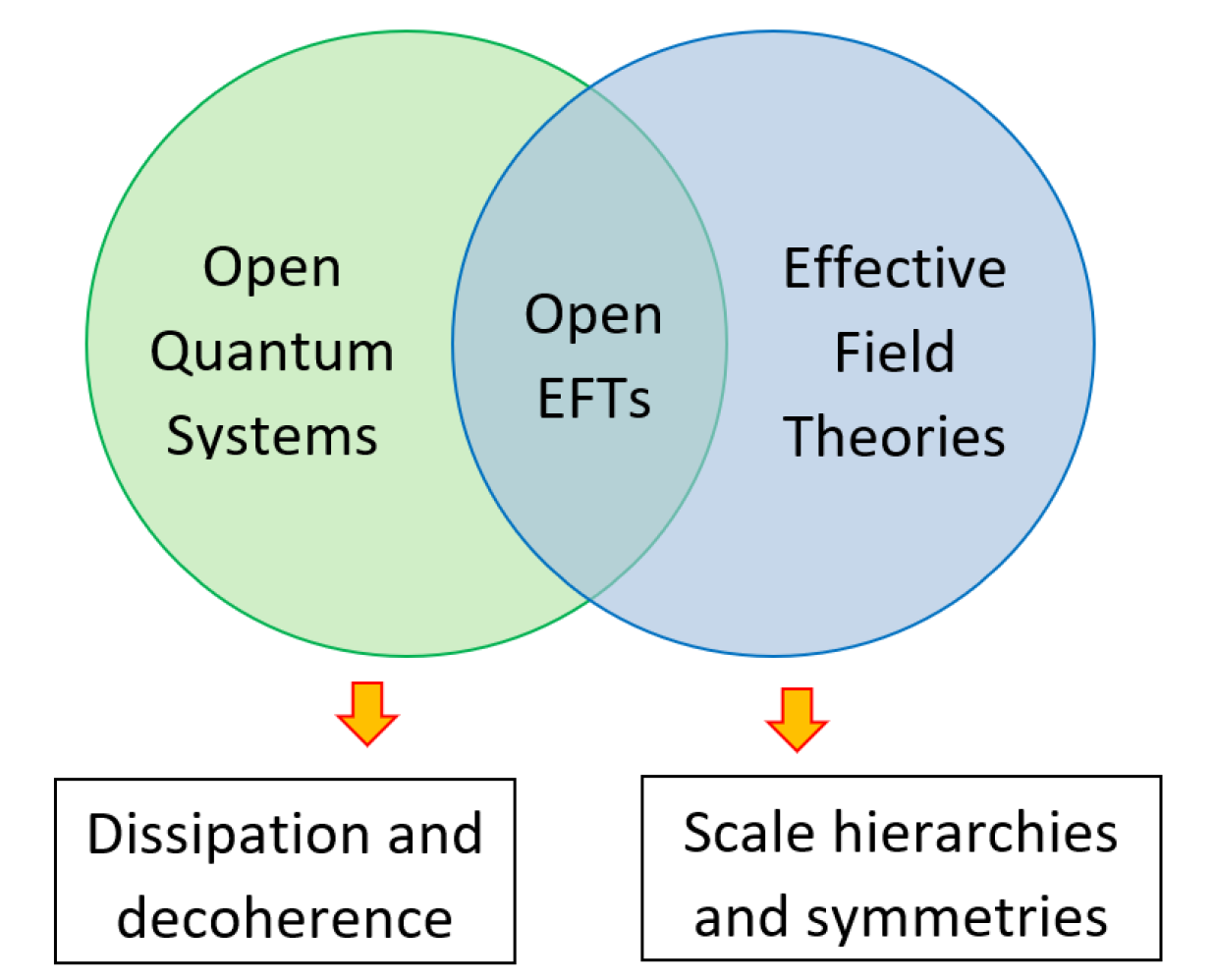} 
	\caption{Open EFTs at the crossroad of OQS techniques and EFTs approaches.}
	\label{fig:Wren}
\end{wrapfigure}

We have seen that many generic cosmological models, in the presence of hidden sectors, exhibit energy and information transfers between observed and unobserved degrees of freedom \cite{Colas:2022kfu}. 

These effects cannot be captured by a local effective action which only accounts for the Hamiltonian evolution of the system \cite{Colas:2022hlq}. We then need to find ways to incorporate dissipative and diffusive dynamics in our EFT dictionnary. Open EFTs precisely aim at achieving this task. They import from Open Quantum Systems (OQS) the knowledge of energy and information exchanges (dissipation and decoherence) while exploiting scale hierarchies and symmetries through EFT techniques. In this way, their goal consists of extending the range of accessible EFTs describing cosmological evolutions, from usual constructions such as the EFT of inflation \cite{Cheung:2007st} to dissipative and diffusive systems such as stochastic inflation \cite{Starobinsky:1986fx} or warm inflation \cite{Berera:1995ie}. In this proceedings, we briefly highlight a recently proposed bottom-up construction \cite{Salcedo:2024smn}. 

\section{The Open Effective Field Theory of Inflation}\label{sec:bottomup}

In \cite{Salcedo:2024smn}, we constructed a bottom-up open EFT for inflation, building on past work of \cite{LopezNacir:2011kk} and \cite{Hongo:2018ant}. Let us consider the Schwinger-Keldysh generating functional of cosmological correlators
\begin{align}
	\mathcal{Z}\left[J_r, J_a \right] = \int \mathcal{D} \pi_r \mathcal{D} \pi_a e^{iS_{\mathrm{eff}}\left[  \pi_r,  \pi_a\right]} e^{i \int \dd^4 x \sqrt{-g}\left(J_r \pi_r + J_a \pi_a\right)},
\end{align}
where $ \pi_r$ is the pseudo-Goldstone boson of the time-translation symmetry breaking induced by the inflaton background. By functional derivative with respect to the source $J_r$, we extract the stastistics of $ \pi_r$ which directly relates to the curvature perturbations $\zeta$. The second field $ \pi_a$ is known as the \textit{advanced} field which is instrumental in describing non-equilibrium processes \cite{kamenev_2011}. 

\paragraph{Construction rules} Our goal is to spell the rules obeyed by the \textit{open effective functional} $S_{\mathrm{eff}}\left[  \pi_r,  \pi_a\right]$ that weighs the in-in path integral. In  \cite{Salcedo:2024smn}, we constructed the most generic $S_{\mathrm{eff}}\left[  \pi_r,  \pi_r\right]$ compatible with (i) \textit{unitarity of the UV theory}; (ii) \textit{the spontaneous symmetry breaking of time-translations}; and (iii) \textit{locality} in time and space of the open effective theory. Condition (i) provides a set of non-perturbative relations known as \textit{non-equilibrium constraints} \cite{Liu:2018kfw} 
\begin{align}
	S_{\mathrm{eff}} \left[ \pi_r, \pi_a = 0\right] &= 0\,, \label{eq:normintro} \\
	S_{\mathrm{eff}} \left[ \pi_r, \pi_a\right] &= - 	S^*_{\mathrm{eff}} \left[ \pi_r,- \pi_a\right]\,, \label{eq:hermintro} \\
	\Ima S_{\mathrm{eff}} \left[ \pi_r, \pi_a\right] &\geq 0. \label{eq:posintro}
\end{align} 
The symmetry requirement further restricts the available dynamics. Breaking the time-translation symmetry leads to two Stueckelberg fields in the unitary case, one for each branch of the in-in path integral. Dissipative and diffusive effects further break the symmetry group explicitly to its diagonal subgroup, such that only $ \pi_{r}$ transforms non-linearly under time-translations and boosts \cite{Hongo:2018ant}. Explicitly, for $\epsilon \in \mathbb{R}$ and $\Lambda^\mu_{~\nu} \in \mathrm{SO}(1,3)$
\begin{align}\label{eq:LorentzRsymintro}
	 \pi_{r}(t,\bmx) \rightarrow  \pi'_{r}(t,\bmx) &=  \pi_{r}\left(\Lambda^0_{~\mu}x^\mu+\epsilon,\Lambda^i_{~\mu}x^\mu \right)+\epsilon+\Lambda^0_{~\mu}x^\mu - t\,, \\
	 \pi_{a}(t,\bmx) \rightarrow  \pi'_{a}(t,\bmx) &=  \pi_{a}\left(\Lambda^0_{~\mu}x^\mu+\epsilon,\Lambda^i_{~\mu}x^\mu \right).
\end{align}
Finally, locality in time and space ensures the existence of an IR-stable power counting scheme that can be truncated to the desired level of accuracy.

\paragraph{Open effective functional} We work in de Sitter space in the \textit{decoupling limit} \cite{Cheung:2007st} with at most one derivative per field. Under the above construction and at leading order in the slow-roll expansion, the most generic open effective functional reads at the quadratic order 
\begin{align}\label{eq:canonormintro} 
	&\quad S_{\mathrm{eff}}^{(2)} = \int \dd^4 x \Big\{ a^2 \pi'_{r} \pi'_{a} - c_{s}^{2} a^2 \partial_i \pir  \partial^i \pia  \\
	-&  a^3 \gamma \pi'_{r} \pia + i \left[\beta_{1} a^4 \pia^2 - \left(\beta_2 - \beta_4\right) a^2\pia^{\prime 2} + \beta_2 a^2 \left(\partial_i \pia \right)^2\right] \Big\}, \nonumber
\end{align}
and the cubic order
\begin{align}\label{eq:canonormcubintro}
	S_{\mathrm{eff}}^{(3)} =   \frac{1}{f_\pi^2} \int & \dd^4 x \Big\{\Big[4 \alpha_2 -  \frac{3}{2} (c^2_s-1) \Big]  a \pir^{\prime2} \pi'_{a} +\frac{1}{2} (c^2_s-1) a \left[\left(\partial_i \pir \right)^2 \pi'_a + 2 \pi'_{r}	\partial_i \pir  \partial^i \pia \right] \\
	& \qquad \quad + \left(4 \gamma_2 - \frac{\gamma}{2}\right)  a^2\pir^{\prime2} \pia	+ \frac{\gamma}{2} a^2	\left(\partial_i \pir \right)^2 \pia \nonumber \Big. \\
	&+ i \Big[\left(2\beta_7-\beta_3\right) a^2 \pi'_{r} \pi'_{a} \pia + \beta_3 a^2 \partial_i \pir  \partial^i \pia \pia + 2(\beta_4+ \beta_6 - \beta_8) a \pi'_{r} \pia^{\prime2} \Big. \nonumber \\
	& \qquad \quad - 2 \beta_4  a \partial_i \pir  \partial^i \pia \pi'_{a} - 2\beta_5 a^3 \pir^{\prime} \pia^2  - 2 \beta_6 a \pir^{\prime} (\partial_i\pia)^2 \Big]  \Big. \nonumber\\
	&+\delta_1 a^4 \pia^3 + (\delta_5- \delta_2) a^2 \pia^{\prime2} \pia  + \delta_2 a^2 (\partial_i \pia)^2 \pia - \delta_4 a  (\partial_i \pia)^2 \pi'_a + (\delta_4-\delta_6) a \pia^{\prime3} \Big\}.  \nonumber
\end{align}
Primes denote time derivatives with respect to the conformal time $\eta = - 1/(aH)$ where $a$ is the scale factor and $H$ the Hubble parameter. The EFT coefficients $f_\pi$, $c_s$, $\alpha_i$, $\beta_i$, $\gamma_i$ and $\delta_i$ are free parameters one can fit to the data (background and fluctuations). While the standard effective field theory of inflation \cite{Cheung:2007st} is recovered in the unitary limit, the above open effective functional also captures non-unitary effects such as dissipation and diffusion of the pseudo-Goldstone boson in an unknown surrounding environment. For instance, the first line of \Eq{eq:canonormintro} corresponds to the usual unitary dynamics with the kinetic term and an effective speed of sound $c_s$, whereas the second line of \Eq{eq:canonormintro} captures dissipation (controlled by $\gamma$) and noise fluctuations (controlled by $\beta_{1}$, $\beta_2$ and $\beta_4$). Just as in the usual EFToI \cite{Cheung:2007st}, non-linearly realised boosts relate non-unitary operators at different orders such as the dissipation parameter $\gamma [- a \pir^{\prime}  -\pir^{\prime2}/2 +\left(\partial_i \pir \right)^2/2] \pia $ \cite{LopezNacir:2011kk}.

\paragraph{The power spectrum} The open effective field theory of inflation provides theoretical predictions for standard cosmological observables such as the power spectrum and the bispectrum. Symmetry requirements ensure the existence of a nearly scale invariant power spectrum
\begin{align}
	\Delta^2_\zeta(k) \equiv \frac{k^3}{2\pi^2}P_\zeta(k) \qquad \mathrm{with} \qquad \langle \zeta_\bmk \zeta_{-\bmk}\rangle = (2\pi)^3 \delta(\bmk + \bmk') P_\zeta(k).
\end{align}
Considering the first noise term of \Eq{eq:canonormintro}, which is controlled by $\beta_1$, we obtained the dissipative power spectrum
\begin{align}
	\Delta^2_\zeta(k) &= \frac{\beta_1}{H^2}   \frac{H^4}{f_\pi^4}  2^{1+\frac{\gamma}{H}} \frac{\Gamma\left(\frac{1}{2}+ \frac{\gamma}{2H}\right)\Gamma\left(\frac{3}{2}+ \frac{\gamma}{2H}\right)^2}{\Gamma\left(1+ \frac{\gamma}{2H}\right)\Gamma\left(\frac{5}{2}+ \frac{\gamma}{H}\right)} \propto  \begin{dcases}
		\frac{\beta_1}{H^2} \frac{H^4}{f_\pi^4}   \sqrt{\frac{H}{\gamma}}\left[1
		 + \mathcal{O}\left(\frac{H}{\gamma}\right)\right], & \gamma \gg H,\\
		\frac{\beta_1}{H^2} \frac{H^4}{f_\pi^4} \left[1 + \mathcal{O}\left(\frac{\gamma}{H}\right)\right], & \gamma \ll H.
	\end{dcases}
\end{align}
The observational constraint $\Delta^2_\zeta = 10^{-9}$ can be easily obeyed by imposing hierarchies between the various scales of the problem. One can further impose thermal equilibrium of the surrounding environment. Then, the power spectrum scales in the large dissipation regime as
\begin{align}
	\Delta^2_\zeta \propto \frac{T}{H}\frac{H^4}{f_\pi^4}\, \sqrt{\frac{\gamma}{H}},
\end{align}
in terms of the environment temperature $T$. This reproduces the warm inflation results \cite{Berera:1995ie}. Analytical results for the two other noise directions $\pia^{\prime 2}$ and $\left(\partial_i \pia \right)^2$ can also be found in \cite{Salcedo:2024smn}. 

\begin{figure}[tbp]
	\begin{minipage}{6in}
		\centering
		\raisebox{-0.5\height}{\includegraphics[width=.48\textwidth]{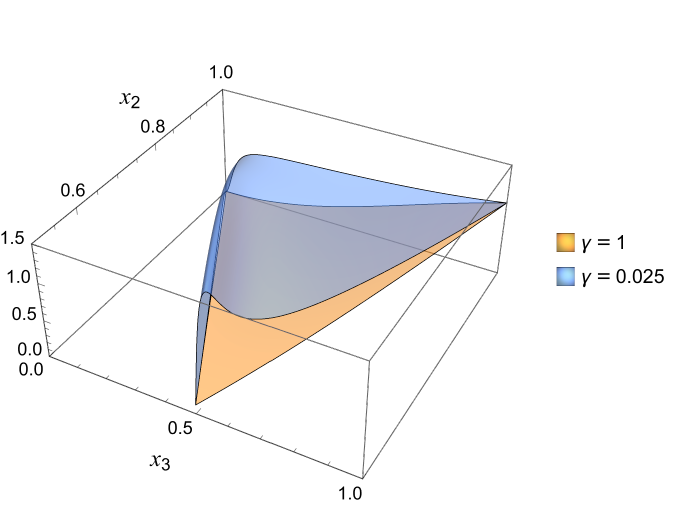}}
		\raisebox{-0.5\height}{\includegraphics[width=.48\textwidth]{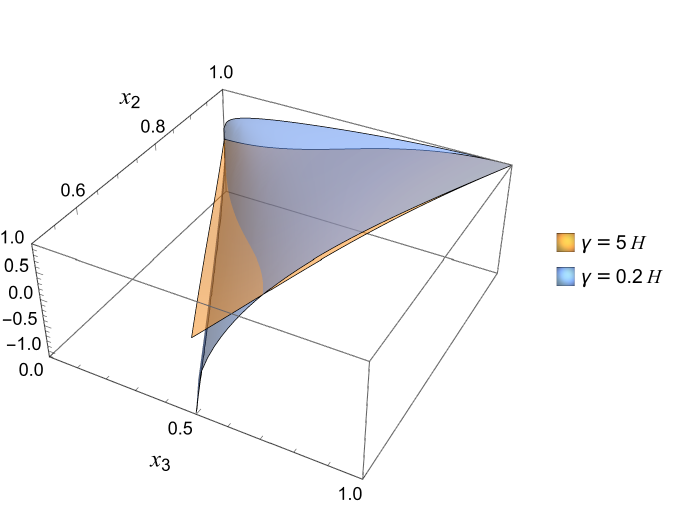}}
	\end{minipage}
	\caption{\label{fig:foldedintro} Shape function $S(x_2,x_3) \equiv (x_2 x_3)^2 [B(k_1, x_2 k_1, x_3 k_1)/B(k_1,k_1,k_1)]$ for the contact bispectrum generated by the operator $\pi_a^3$ in Minkowski (\textit{left}) and in de Sitter (\textit{right}). \textit{Left}: the bispectrum is given by \Eq{eq:genintro} with singularities controlled by $\mathrm{Sing}_\gamma$ in \Eq{eq:singdissipintro}. The singularity is resolved such that the bispectrum remains finite for any physical configuration. We observe the equilateral (\textit{orange}) to folded (\textit{blue}) transition of the shape function as the dissipation parameter $\gamma$ decreases. There is an enhancement of the signal close to the isofolded configuration in the low dissipation regime. \textit{Right}: the qualitive features of the signal remain the same in de Sitter. At large dissipation (\textit{orange}), the signal peaks in the equilateral configuration. At small dissipation (\textit{blue}), it is enhanced near the folded region. Two aspects are confirmed by analytical arguments but are not clearly visible in the plot: (i) dissipation regulates the folded divergence and (ii) consistency relations still hold in the squeezed limit $x_3 \ll x_2 = 1$.}
\end{figure}

\paragraph{The bispectrum} Beyond the Gaussian case, one can use the same treatment as in the standard in-in formalism \cite{Chen:2017ryl} to account for non-linear interactions. The bispectrum
\begin{align}
	\langle \zeta_{\bmk_1}  \zeta_{\bmk_2}  \zeta_{\bmk_3} \rangle  = - \frac{H^3}{f_\pi^6}	\langle \pi_{\bmk_1}  \pi_{\bmk_2}  \pi_{\bmk_3} \rangle \equiv (2\pi)^3 \delta(\bmk_1 + \bmk_2 + \bmk_3) B(k_1,k_2,k_3).
\end{align} 
is generated by the cubic operators in \Eq{eq:canonormcubintro}, both in flat space and in de Sitter. The flat-space results are instructive as computations can easily be carried out analytically. The generic structure of the contact bispectrum is given by 
\begin{align}\label{eq:genintro}
	B(k_1,k_2,k_3) =f(\mathrm{EFT}) \frac{  \mathrm{Poly}_n\left(e_1^\gamma, e_2^\gamma,e_3^\gamma \right)}{ \mathrm{Sing}_\gamma} \,,
\end{align}
where $f(\mathrm{EFT})$ is a rational function of the EFT coefficients (and possibly the kinematics for spatial derivative interactions), and $\mathrm{Poly}_n$ are elementary symmetric polynomials of the energy variables
\begin{align}\label{eq:varintro}
	e_1^\gamma = E_1^{\gamma} + E_2^{\gamma} + E_3^{\gamma}, \quad e_2^\gamma = E_1^{\gamma} E_2^{\gamma} + E_2^{\gamma}  E_3^{\gamma} + E_1^{\gamma} E_3^{\gamma} \quad  e_3^\gamma = E_1^{\gamma} E_2^{\gamma} E_3^{\gamma}\,,
\end{align}
where we used the dispersion relation appropriate for this dissipative system, namely 
\begin{align}
	E_k^{\gamma} \equiv \sqrt{c_s^2 k^2 - \gamma^2/4}\,.   
\end{align}
Moreover, $\mathrm{Sing}_\gamma$ is a place holder for the singularity structure
\begin{align}\label{eq:singdissipintro}
	\mathrm{Sing}_\gamma =& \left| E_1^{\gamma} + E_2^{\gamma} + E_3^{\gamma} + \frac{3}{2}i \gamma\right|^2 \left| -E_1^{\gamma} + E_2^{\gamma} + E_3^{\gamma} + \frac{3}{2}i \gamma\right|^2 \nonumber\\
	&\times \left| E_1^{\gamma} - E_2^{\gamma} + E_3^{\gamma} + \frac{3}{2}i \gamma\right|^2 \left| E_1^{\gamma} + E_2^{\gamma} - E_3^{\gamma} + \frac{3}{2}i \gamma\right|^2.
\end{align}
This singularity structure captures most of the specificities of the non-unitary dynamics. Physically, it represents $3 \leftrightarrow 0$ (all pluses) and $2 \leftrightarrow 1$ (mixed signs) interactions mediated by the real particles present in the environment. Fluctuations alone would generate \textit{folded singularities} because the state of the system differs from the Bunch Davies vacuum. On the other hand, dissipation induces a finite memory of the past. This regularises the folded divergences \cite{LopezNacir:2011kk}, by effectively moving the folded pole into complex kinematics. The singularity is not located in the physical plane and the bispectrum remains finite over the whole dynamical range. This is illustrated in the \textit{left} panel of \Fig{fig:foldedintro} where we observe an enhanced but finite signal near to folded region.  
The singularity structure $\mathrm{Sing}_\gamma$ exhibits two different behaviours depending on the magnitude of the dissipation coefficient $\gamma$: 
\begin{itemize}
	\item In the strong dissipation regime, the $ 3 i \gamma/2$ term of \Eq{eq:singdissipintro} always dominates and the signal peaks in the \textit{equilateral} shape where $k_1 \simeq k_3 \simeq k_3$ (\textit{orange region} in \Fig{fig:foldedintro}). 
	\item In the small dissipation regime, $\mathrm{Sing}_\gamma$ can become small in the \textit{folded} region where $k_2 + k_3 \simeq k_1$ and the signal predominantly peaks near $k_2 \simeq k_3 \simeq k_1/2$ (\textit{blue region} in \Fig{fig:foldedintro}).
\end{itemize} 
Beyond the flat space case, analytical results are hard to reach in full generality and we mostly rely on numerical results. Just as in flat space, different behaviours emerge in the large ($\gamma \gg H$) and small ($\gamma \ll H$) dissipation regime, see the \textit{right} panel of \Fig{fig:foldedintro}. For large dissipation the signal peaks in the equilateral configuration, as already noted in \cite{LopezNacir:2011kk}; for small dissipation the signal reaches an extremum near the folded region. At first sigh, this smoking gun of open dynamics might seem to be degenerate with other classes of models, which also lead to signal in the folded triangles, such as non-Bunch Davies initial states \cite{Holman:2007na}. A crucial difference is that dissipation regulates the divergence by smoothing the peak and displacing it from the edge of the triangular configurations, leading to finite values of the bispectrum for any physical configuration. In particular, it implies no divergence in the squeezed limit of the bispectrum $k_1 \simeq k_2 \gg k_3$. Small values of $\gamma/H$ may eventually lead to an intermediate peak due to the regularised folded singularity, but consistency relations hold \cite{Maldacena:2002vr} and the squeezed limit goes to zero due to the symmetries of the theory.

\paragraph{UV-models and matching} The open effective field theory of inflation captures all single-clock models that display a local and possibly dissipative dynamics. One such explicit UV-model was recently studied in \cite{Creminelli:2023aly}. We shown in \cite{Salcedo:2024smn} that the above formalism provides an accurate low-energy effective description of its dynamics. The model in question contains, in addition to the inflaton field $\phi$, a massive scalar field $\chi$ with a softly-broken $U(1)$ symmetry
\begin{align}\label{eq:actionmatchintro}
	S= \int \dd^4 x &\sqrt{- g} \bigg[ \frac{1}{2} M_{\mathrm{Pl}}^2  R - \frac{1}{2} \left(\partial \phi \right)^2 - V(\phi) - \left|\partial \chi \right|^2 + M^2 \left| \chi \right|^2 \nonumber \\
	- &\frac{\partial_\mu \phi}{f}\left( \chi \partial^\mu \chi^* - \chi^* \partial^\mu \chi\right) - \frac{1}{2}m^2\left(\chi^2 + \chi^{*2}\right)\bigg].
\end{align}
This model exhibits a narrow instability band in the sub-Hubble regime, during which particle production occurs. It turns out that the non-linear Langevin equation 
\begin{align}\label{eq:Paolointro}
	\pi'' + \left( 2H + \gamma\right)a\pi' - \partial_i^2 \pi &\simeq \frac{\gamma}{2\rho f} \left[ \left( \partial_i \pi\right)^2 - 2 \pi \xi \pi^{\prime2}\right] - \frac{a^2m^2}{f} \left(1+2\pi\xi \frac{\pi'}{a \rho f} \right) \delta \mathcal{O}_S\,,
\end{align}
used in \cite{Creminelli:2023aly} is completely equivalent to the open effective field theory of inflation, which for this UV model reduces to
\begin{align}\label{eq:Seffmatchintro}
	\quad S_{\mathrm{eff}} &= \int \dd^4 x \Big[ a^2 \pi'_{r} \pi'_{a} - c_{s}^{2} a^2 \partial_i \pir  \partial^i \pia   -  a^3 \gamma \pi'_{r} \pia + i \beta_1 a^4 \pia^2 \nonumber \\
	+ &\frac{ \left(8\gamma_2 - \gamma\right)}{2f^{2}_{\pi}}  a^2\pir^{\prime2} \pia	+ \frac{\gamma}{2f_{\pi}^{2}} a^2	\left(\partial_i \pir \right)^2 \pia - 2i\frac{\beta_{5}}{f_\pi^2} a^3 \pir^{\prime} \pia^2 +\frac{\delta_1}{f_\pi^2} a^4 \pia^3  \Big]. 
\end{align}

\section{Conclusion}

There exist situations of cosmological interest which require the inclusion of dissipative and diffusive effects to encode corrections from unknown physics onto observable cosmology. Open EFTs aim at including these effects in a systematic manner. In this proceeding, we briefly presented a recent bottom-up construction \cite{Salcedo:2024smn} which extends the EFT dictionnary to account for the rich phenomenology of local dissipative models of inflation. 

There is a lot of work ahead to further extend this research program. The ability of open EFTs to go beyond perturbation theory by implementing non-perturbative resummations has been highlighted in a variety of top-down models \cite{Burgess:2024eng}. We now have to systematize these constructions to strenghen the early universe predictions. Open EFTs allow us to investigate the quantum information properties of inflation and alternatives \cite{Colas:2024xjy}. It might be a way forward extending flat space techniques to cosmology despite the lack of generally defined S-matrix \cite{Aoude:2024xpx}. On the observational side, signatures of dissipative effects in CMB and LSS data may place bounds on the parameters appearing in the open EFT of inflation. At last, applications to late-time cosmology are direct extensions of this work. 

If we want to understand the rules of EFTs in curved spacetime, we need to seriously consider the specificities of cosmological dynamics where perturbations are created out of the vacuum, energy is not conserved, length and time scales are dynamical, and information is redistributed globally. Open EFTs is a modest step into this direction. \\

\section*{Acknowledgments}

T.C. warmly thanks the organizers of the 2024 Cosmology session of the 58th Rencontres de Moriond and the participants for the stimulating discussions. This work has been supported by STFC consolidated grant ST/X001113/1, ST/T000791/1, ST/T000694/1 and ST/X000664/1.

\section*{References}

\bibliography{biblio}

\end{document}